\documentstyle[preprint,aps]{revtex}
\begin{document}
\title{Periodic Bounce for Nucleation Rate at Finite Temperature in Minisuperspace Models}
\author{J.-Q. Liang$^{1,2}$, H. J. W. M\"{u}ller-Kirsten$^{1}$\footnote{e-mail: mueller1@physik.uni-kl.de}\footnote{corresponding author}, Y.-B. Zhang$^{1,2}$, 
 A. V. Shurgaia$^1$, S.-P. Kou$^3$, D.K. Park$^{1,4}$}
\address{1. Department of Physics, University of Kaiserslautern, D-67653 Kaiserslautern, Germany\\
2. Department of Physics and Institute of Theoretical Physics, Shanxi University, Taiyuan, Shanxi 030006, China\\
3. Institute of Physics and Center for Condensed Matter Physics, Chinese Academy of Sciences, Beijing 100080, China\\
4. Department of Physics, Kyungnam University, Masan, 631-701, Korea}
\maketitle

\begin{abstract}
The periodic bounce configurations responsible for quantum tunneling are obtained explicitly  and are extended to the finite energy case for minisuperspace 
models of the Universe. As a common feature of the tunneling models at finite energy considered here we observe that the period of the bounce  increases with 
energy monotonically. The periodic bounces do not have bifurcations and  make no contribution to the nucleation rate except the one with zero
energy. The 
sharp first order phase transition from quantum tunneling to thermal activation is verified with the general criterions.
\end{abstract}

\section{Introduction}
\label{sec:I}

Quantum tunneling at finite energy and temperature, the so-called thermally assisted tunneling, has attracted considerable attention
recently in the study of the crossover from the quantum tunneling domain to the thermal activation (hopping) region. The instanton method plays a central role in these studies. The probability
of tunneling at zero temperature can be obtained from a micro-cannonical ensemble and has a path integral representation\cite{Gildener}. In the one loop approximation the probability is $P=Ae^{-S}$ 
where the preexponential factor A arises from Gaussian functional integration over small fluctuations around the instanton solution and  S is the Euclidean action of an instanton with zero
energy.\\  
There are two kinds of tunneling, one of which is tunneling between degenerate vacua induced by instantons which are stable Euclidean field solutions with nontrivial topological charge.
 The instanton can be viewed as an extended particle existing in the barrier interpolating between degenerate vacua\cite{Gildener}. A (vacuum) bounce is, however, an unstable solution of a Euclidean field equation
 with zero topological charge and was well known already decades ago\cite{Coleman,Langer}. The initial and end points of a (vacuum) bounce both terminate on a metastable ground state or false vacuum.
 The tunneling induced by such a bounce results in the decay of the false vacuum\cite{Liang1}.\\
Quantum tunneling at finite temperature\cite{Linde1} T is dominated by periodic instantons (bounces) which are periodic solutions of the Euclidean equation of motion with finite energy
E \cite{Khlebnikov,Liang2} and in the
 semi-classical limit
the path integral is expected to be saturated by a single periodic instanton. With exponential accuracy the tunneling probability  P(E) at a given energy E reduces to
\begin{equation}
\label{1}
P(E)\sim e^{-W(E)}=e^{-S(\beta )-E\beta }
\end{equation}
The period $\beta$ is related to the energy E in the standard way $E=\frac{\partial S}{\partial\beta}$ and $S(\beta )$ is the action of the periodic instanton (bounce) per period. Such
periodic instantons (bounces) smoothly interpolate between the  zero temperature instantons (bounces) and the static solution named sphaleron sitting at
the top of the potential barrier. The sphaleron is  responsible for thermal hopping. Peculiarly the study of explicit periodic instantons and their stability began only about
ten years ago\cite{Manton,Liang3}. \\
With increasing temperature thermal hopping becomes more and more important and beyond some critical or crossover temperature $T_c$ becomes the decisive mechanism. In
the context of quantum mechanics it has been demonstrated that the transition from the thermal to the quantum regime can be considered as a phase transition which is
of second-order with certain assumptions about the shape of the potential\cite{Affleck}. Later it was shown that the situation is not generic and that the crossover
 from the thermal to the quantum regime can quite generally be like that of a first-order phase transition\cite{Chudnovsky1}. The sharp first-order transition has been
 confirmed theoretically in several spin tunneling systems\cite{Chudnovsky2,Liang4,Lee} and  triggered active reaserch in various fields in connection with tunneling. In
the context of field theory not much work has been done toward the study of periodic instantons. Recently there were interesting investigations to show that the crossover
from the quantum to the thermal regime in the vacuum decay with $\phi^4$ models is essentially a first-order phase transition in the thin  wall limit\cite{Garriga,Ferrera}.
 It is therefore also a challenging problem to study the crossover from quantum tunneling to thermal activation in the context of 
cosmology\cite{Linde2}. Here we follow the recent model investigations of the creation of Universe in the context of the so-called minisuperspace models\cite{Vilenkin}, 
and extend the study of tunneling to finite energy and temperature.\\
The characteristic way in which phase transitions appear in quantum mechanical tunneling processes has been worked out in ref.[11]. In context of field theory the crossover
behaviour has also been explained in a more transparent manner\cite{Kuznetsov}. A sharp first order transition is shown to appear as a bifurcation in the plot of the instanton
action S versus period $\tau (E)$. The criterion for a first order transition can be obtained by studying the Euclidean time period in the neighbourhood of the sphaleron
as advocated in ref.[20]. If the period $\beta (E\rightarrow U_{0})$ of the periodic instanton (bounce) close to the barrier peak can be found, a sufficiet condition to have the
first order transition is seen to be $\beta (E\rightarrow U_{0})-\beta_{s} < 0$ or $\omega^{2} > \omega^{2}_{s}$, where $U_{0}$ denotes the barrier height and $\beta_{s}$ is
the period of small oscillation around the sphaleron. $\omega $ and $ \omega _{s}$ are the corresponding frequencies. The frequency of the spaleron $\omega _{s}$ is nothing but the frequency of small
oscillaton in the bottom of the inverted potential well. A practically useful formula for the criterion of the first order transition is given in ref.[21] and the winding number transition
in $O(3) \sigma$ model with and withoult Skyrme term has been successfully analyzed with the crirerion\cite{Park}. In the following the crossover
behaviour in the minisuperspace model is investigated in terms of the general criterion and we also explain the physics underlying the crossover which may shed
light on understanding the time evolution of the Universe in the model. In Sec. 2 the quantum tunneling at zero energy is briefly reviewed. We emphasize that the 
bounce starts and ends on the metastable ground state which corresponds to a static solution of the field equation with zero radius and is therefore meaningful for the 
decay of the false vacuum. As a prototype model of the creation of the Universe at finite temperature we discuss the similar process of bubble nucleation in Sec. 3. The 
crossover of the nucleation rate from the quantum to the classical regime is studied in terms of the general criterions for first-order phase transitions. In Sec. 4 we 
apply a similar approach to the cosmological minisuperspace model.

\section{The Periodic Bounce and quantum tunneling at zero energy}
\label{sec:II}

Contemporary cosmological models are based on the idea that the Universe is pretty much the same everywhere $-$ an idea sometimes known as the Copernican principle which is 
related to two more mathematically precise properties that the manifold might have: isotropy and homogeneity. We begin with the simplest minisuperspace model of the 
Universe\cite{Vilenkin} defined by the action:
\begin{equation}
  \label{2}
  S=\int d^{1+N}x\sqrt{-g}\left[ \frac{R}{16 \pi G_N}-\rho_v \right]
\end{equation}
where $\rho_v>0$ is a constant vacuum energy according to Ref.\cite{Vilenkin} and thus plays the role of the cosmological constant in eq. (\ref{2}) and makes the space 
de Sitter or anti-de Sitter. The spacetime to be considered is ${\bf R}\times \Sigma$ where ${\bf R}$ represents the time direction and $\Sigma$ is a homogeneous and 
isotropic $N-$manifold with $N=2$ or $3$. The Universe is also assumed to be closed. We therefore have:
\begin{equation}
  \label{3}
  ds^2=-dt^2+\xi ^2(t)d\Omega_N^2
\end{equation}
which is just the Robertson$-$Walker metric of the closed case. The function $\xi(t)$ is known as the scale factor which tells us ``how big'' the spacetime slice $\Sigma$ is 
at time $t$. $d\Omega_N^2$ is the metric on a unit $N-$sphere. Substituting the metric eq.(\ref{3}) into eq.(\ref{2}) we obtain the Lagrangian
\begin{equation}
  \label{4}
  {\cal L}=-S_N\xi^{N-2}\left[\frac{N(N-1)}{16 \pi G_N}(1-{\dot \xi}^2)-\xi^2\rho_v \right]
\end{equation}
where
\begin{equation}
  \label{5}
  S_N=\frac{2 \pi^{\frac{N+1}{2}}}{\Gamma(\frac{N+1}{2})}
\end{equation}
is the surface of the unit $N-$sphere. It is easy to see that only for $N=3$ is $\xi=0$ a static solution of the equation of motion and thus can serve as the metastable ground state or false vacuum. The Lagrangian for $N=3$ can be written
\begin{equation}
  \label{6}
  {\cal L}=\frac 12 M(\xi) {\dot \xi}^2-V(\xi)
\end{equation}
where $M(\xi)=m_0\xi$ is the position dependent mass with $m_0=\frac{3 \pi}{2 G}$. The potential
\begin{equation}
  \label{7}
  V(\xi)=\frac{m_0}{2}\xi-\eta \xi^3
\end{equation}
is shown in Fig. 1 where $\eta=2 \pi^2 \rho_v$. The classical solution of the equation of motion in real time is\cite{Vilenkin}
\begin{equation}
  \label{8}
  \xi(t)=\frac 1\Omega \cosh \Omega t, \qquad \Omega=\sqrt{\frac{2 \eta}{m_0}}
\end{equation}
which shows that the space is the de Sitter space expanding at $t>0$ from $\xi(t=0)=\frac 1\Omega$. $\xi=0$ is an additional static solution  with energy $E=0$. The bounce
 configuration is obtained from the Euclidean equation of motion by the Wick rotation $\tau=it$  under the barrier and is seen to be
\begin{equation}
  \label{9}
  \xi_b(\tau)=\frac 1\Omega \cos \Omega \tau, \qquad
  \left[ -\frac{\pi}2 \leq \Omega \tau \leq\frac \pi 2\right]{\rm mod} 2 \pi
\end{equation}
We see that the bounce is a periodic solution. The trajectory of this periodic bounce for one period is shown in Fig. 1a. The bounce starts from the false vacuum ($\xi=0$) 
at imaginary time $\tau=-\frac{\pi}{2 \Omega}$ and reaches the turning point $\xi=\frac 1 \Omega$ at time $\tau=0$ and then bounces back to the false vacuum at
 $\tau= \frac{\pi}{2 \Omega}$. The period of the bounce is
\begin{equation}
  \label{10}
  \beta=\frac \pi\Omega.
\end{equation}
The Universe can then be considered to be created spontaneously from ``nothing'' ($\xi=0$) and to tunnel through the barrier (Fig. 1) into the de Sitter space. The 
tunneling rate or decay rate out of the false vacuum can be evaluated in terms of the action of the bounce and is given by
\begin{equation}
  \label{11}
  P(E=0) \sim e^{-W_b}
\end{equation}
where
\begin{equation}
  \label{12}
  W_b=\int_{\xi(\tau=-\frac{\pi}{2 \Omega})=0}^{\xi(\tau=\frac{\pi}{2 \Omega})=0}p_b(\xi)d\xi=\frac{3}{8 G^2 \rho_v}
\end{equation}
Here $p_b$ denotes the momentum of the bounce and is as usual evaluated from the Euclidean version ${\cal L}_E$ of the Lagrangian eq.(\ref {6}), i.e. 
\begin{equation}
  \label{13}
  p_b=\frac{\partial {\cal L}_E}{\partial \dot \xi}|_{\xi=\xi_b}=m_0\dot \xi_b
\end{equation}
It may be noted that the bounce here is periodic even though the energy of the false vacuum is taken to be zero. This is quite unlike the usual case of the 
bounce at zero energy as, for example, in the case of the well studied bounce of the inverted double-well potential\cite{Liang1} where the period of the bounce tends to infinity. The 
periodic bounce with finite period exists only at finite energy\cite{Liang2}. \\

We now turn to bubble nucleation in the thin wall case as a comparison. When a field configuration is trapped in a metastable state, bubbles of the true vacuum state 
nucleate in the surrounding false vacuum and begin to grow spherically. The process of bubble nucleation is in many ways analogous to the nucleation of the Universe. 
Under a number of simplifying assumptions the nucleating bubble can be adequately described by a minisuperspace model with a single degree of freedom\cite{Vilenkin}, the 
bubble radius $r(t)$. The Lagrangian of $1+N$ dimensions is
\begin{equation}
  \label{14}
  {\cal L}=-S_{N-1}\left[ r^{N-1}(1-\dot r^2)^{\frac 12} \sigma -\frac{\epsilon}{N}r^N\right]
\end{equation}
Here $\sigma$ is the tension of the wall, $N=2,3$, and $\epsilon$ denotes the difference in the vacuum energy on both sides of the wall. The canonical momentum conjugate 
to the variable $r$ is
\begin{equation}
  \label{15}
  p=\sigma S_{N-1} \frac{\dot r r^{N-1}}{(1-\dot r^2)^{\frac 12}}
\end{equation}
and the Hamiltonian is
\begin{equation}
  \label{16}
  H=\left[ p^2+\sigma^2S_{N-1}^2r^{2(N-1)}\right]^{\frac 12}-\frac{\epsilon S_{N-1}}{N}r^N
\end{equation}
We then obtain a point particle like Hamiltonian which is the starting point of our considerations. The energy is conserved in the process of bubble nucleation. For zero 
energy the equation $H=0$ can be rewritten as
\begin{equation}
  \label{17}
  p^2+U(r)=0
\end{equation}
with the effective potential
\begin{equation}
  \label{18}
  U(r)=\sigma^2 S_{N-1}^2r^{2(N-1)} \left[ 1-\frac{r^2}{r_0^2}\right]
\end{equation}
where $r_0=\frac{N\sigma}{\epsilon}$. We see that for both $N=2$ and $3$ the vanishing radius $r=0$ is a static solution of the equation of motion 
 with zero energy, namely, the vacuum of our point particle like system. Besides the static solution $r=0$ the solution in real time is
\begin{equation}
  \label{19}
  r(t)=(r_0^2+t^2)^{\frac 12}
\end{equation}
It is shown in Ref. \cite{Vilenkin} that this solution eq. (\ref{19}) is the same de Sitter space eq. (\ref{8})
\begin{equation}
  \label{20}
  r(\tilde t)=r_0 \cosh (\frac{\tilde t}{r_0})
\end{equation}
in the new time coordinate $\tilde t$ with
\begin{equation}
  \label{21}
  t=r_0 \sinh \frac{\tilde t}{r_0}
\end{equation}
In the new imaginary time coordinate $\tilde \tau=i\tilde t$ the imaginary time solution existing in the barrier is just the periodic bounce of eq.(\ref{9}), i.e.
\begin{equation}
  \label{22}
  r_b(\tilde \tau)=r_0 \cos \frac{\tilde \tau}{r_0}, \qquad
  \left[ -\frac{\pi}2 \leq \frac {\tilde \tau}{r_0} \leq\frac \pi 2\right]{\rm mod} 2 \pi
\end{equation}
with the finite period
\begin{equation}
  \label{23}
  \tilde \beta=\pi r_0
\end{equation}
The action of the bounce is
\begin{equation}
  \label{24}
  W_b=\int_{r(\tau=-\frac{\tilde\beta}{2})=0}^{r(\tau=\frac{\tilde\beta}{2})=0}p_bdr=\Biggl \lbrace _{\frac{\pi^2}2 \sigma r_0^3, \qquad for\quad N=3 }^{\frac{4 \pi}{3} \sigma r_0^2, \qquad for \quad N=2}
\end{equation}
We see that the bubble nucleation is indeed similar to the creation of the Universe.

\section{Crossover from quantum tunneling to thermal activation $-$ bubble nucleation}
\label{sec:III}

As a prototype for the nucleation of the Universe we reconsider the temperature dependence of the bubble nucleation rate. However, we study the crossover from the quantum to the classical 
regime in terms of the general criteria for phase transitions\cite{Gorokhov,Muller-Kirsten,Park}. We consider the nucleation process at finite energy $E$. Then energy conservation $H=E$ leads to
\begin{equation}
  \label{25}
  p^2+U(r,E)=0
\end{equation}
with the effective potential (see Fig. 2)
\begin{equation}
  \label{26}
  U(r,E)=\sigma^2 S_{N-1}^2 \left[ r^{2(N-1)} -(\frac{E}{\sigma S_{N-1}}+\frac{r^N}{r_0})^2\right]
\end{equation}
The periodic bounce at finite energy $E$ is an imaginary time solution which exists in the barrier between two turning points $r_\pm$ (as shown in Fig. 2) which are static solutions of the field equation (\ref{25}). The parameters in Fig. 2 are defined by
\begin{equation}
  \label{27}
  r_\pm=\frac{r_0}{2}(1 \pm \sqrt{1-\delta}),\quad \delta=\frac{E}{U_0}, \quad U_0=\frac{\sigma S_1 r_0}{4}
\end{equation}
for $N=2$. The bounce of eq.(\ref{9}) is recovered when the energy reduces to zero, $E=0$. The period of the bounce for $N=2$ is\cite{Garriga}
\begin{equation}
  \label{28}
  \beta(E)=2\left[r_+ {\cal E}(k)+r_-{\cal K}(k) \right]
\end{equation}
where ${\cal K}(k)$ and ${\cal E}(k)$ denote the complete elliptic integrals of the first and second kinds respectively with modulus
\begin{equation}
  \label{29}
  k^2=1-\frac{r_-^2}{r_+^2}
\end{equation}
The period $\beta (E)$ increases monotonically with energy from its minimum value $2 r_0$ at zero energy to the maximum value $\pi r_0$ at energy $E$ reaching the upper bound at $E=U_0$. When the bubble with radius $r_-$ is spontaneously created it may decay through the barrier by quantum tunneling. The tunneling rate is again calculated from the action of the bounce\cite{Garriga} ($N=2$)
\begin{equation}
  \label{30}
  W_{b}=\int_{r_-(\tau=-\frac{\beta}{2})}^{r_-(\tau=\frac{\beta}{2})}p_bdr=\frac{2 \sigma S_1 r_+}{3 r_0}\left[ (r_+^2+r_-^2) {\cal E}(k)-2 r_-^2 {\cal K}(k)\right]
\end{equation}
The shape of the potential barrier varies with energy $E$ as shown in Fig. 2. When the energy reaches the upper bound $U_0$ the two static solutions $r_+,r_-$ join at the top of the barrier and the solution is called the sphaleron
\begin{equation}
  \label{31}
  r_s=\frac {r_0} 2
\end{equation}
which plays an important role in the crossover from quantum tunneling to thermal activation. \\

Our main point here is to investigate the transition from the quantum to the classical regime. The crossover is realized as a phase transition analogous to the Landau
theory. To this end we start from a procedure similar to that in Ref. \cite{Chudnovsky2} where the phase transition in the tunneling rate of a spin system is discussed. 
We expand the bounce action $W_b$ around the sphaleron ($E \rightarrow U_0, \delta \rightarrow 1, k\rightarrow 0$) and use the series expansions of the complete elliptic
 integrals
\begin{eqnarray}
  \label{32}
  {\cal K}(k)=\frac \pi 2 \left[ 1+\frac 14 k^2+\frac 9{64}k^4+\cdots \right]\\
  {\cal E}(k)=\frac \pi 2 \left[ 1-\frac 14 k^2-\frac 3{64}k^4+\cdots \right] \nonumber
\end{eqnarray}
Defining a new parameter $h=1-\frac{E}{U_0}=1-\delta$, the modulus $k$ of the complete elliptic integrals and the turning points $r_\pm$ are expressed in terms of $h$,
\begin{equation}
  \label{33}
  k^2=1-\left(\frac{1-\sqrt{h}}{1+\sqrt{h}} \right)^2, \quad r_\pm=\frac{r_0}2 (1\pm\sqrt{h})
\end{equation}
Substituting the expansion eq. (\ref{32}) into eq. (\ref{30}) the free energy of the bounce, $F=E+TW$, near the sphaleron is then expanded as a power series of $h$
\begin{equation}
  \label{34}
  \frac{F}{U_0}=1+(\theta-1)h-\frac 18 \theta h^2-\frac 1{64}\theta h^3+O(h^4)
\end{equation}
where $\theta=\frac{T}{T_s}$ is the dimensionless temperature with $T_s=\frac{1}{\beta_s}$ and $\beta_s=\pi r_0$ is the period of the sphaleron. The analogy with the 
Landau theory of phase transitions described by $F=a\psi^2+b\psi^4+c\psi^6$ where $\psi$ is the order parameter is obvious. The factor in front of $h$ changes its sign 
at the phase transition temperature $T_s$. The factor in front of $h^2$ has always the negative sign which indicates the first-order phase transition\cite{Chudnovsky2}.  

Recently the phase transition from quantum to classical regime has been studied extensively. A criterion of the first-order phase transition has been formulated for the 
crossover from quantum tunneling to thermal activation\cite{Gorokhov,Muller-Kirsten}. The key point in the procedure is to investigate the quantum fluctuation around the sphaleron. 
The oscillation frequency around the sphaleron can be expanded as a perturbation series\cite{Muller-Kirsten}
\begin{equation}
  \label{35}
  \omega^2=\omega_s^2+\lambda \Delta_1 \omega^2 +\lambda^2\Delta_2\omega^2+\cdots
\end{equation}
where $\omega_s=\frac{2 \pi}{\beta_s}$ is the frequency of the sphaleron and $\lambda$ denotes the perturbation parameter. It is demonstrated\cite{Muller-Kirsten} that the criterion 
for the first-order phase transition $\Delta _2 \omega^2>0$ leads to a useful inequality derived from the Euclidean equation of motion, i.e., the bounce trajectory\cite{Muller-Kirsten}, namely
\begin{equation}
  \label{36}
  V^{(3)}(r_s)(g_1+\frac{g_2}2)+\frac 18 V^{(4)}(r_s)+M^{(1)}(r_s)(g_1+\frac{3g_2}2)+\frac 14 M^{(2)}(r_s)\omega_s^2<0
\end{equation}
where $f^{(n)}(r_s) \equiv \frac{d^nf(r)}{dr^n}|_{r=r_s}$ is defined as the usual n-th partial derivative at the coordinate of the sphaleron, and
\begin{eqnarray*}
  g_1=-\frac{\omega_s^2 M^{(1)}(r_s)+V^{(3)}(r_s)}{4V^{(2)}(r_s)}, \qquad
 g_2=-\frac{3\omega_s^2 M^{(1)}(r_s)+V^{(3)}(r_s)}{4\left[4M(r_s)\omega_s^2+V^{(2)}(r_s)\right]},
\end{eqnarray*}
$M(r)$ is the mass and is generally position dependent (as for example in eq.(\ref{6})). The criterion for the first-order phase transition contains only the information of the sphaleron. It is not necessary to obtain the bounce configuration for the entire region of energy ($E=0$ to $U_0$). We now apply the criterion (\ref{36}) to the problem of bubble nucleation for both $N=2$ and $3$. From the equation of motion (\ref{25}) we derive
\begin{eqnarray*}
  V(r,E)&=&\left[\frac{E}{\sigma S_{N-1}}r^{1-N}+\frac{\epsilon}{N\sigma}r \right]^{-2}-1,\\
  V^{(1)}(x_s,\varepsilon)&=&\frac{2\left[ (N-1)\varepsilon x^{-N}-1\right]}{(\varepsilon x^{1-N}+x)^3}=0,\\
  V^{(2)}(x_s,\varepsilon)&=&-\frac{2(N-1)^3}{N^2[\varepsilon (N-1)]^{\frac 4N}}, \\
  V^{(3)}(x_s,\varepsilon)&=&\frac{2(N-1)^3(N+1)}{N^2[\varepsilon (N-1)]^{\frac 5N}}, \\
  V^{(4)}(x_s,\varepsilon)&=&-\frac{2(N-1)^3(N^2-6N+11)}{N^2[\varepsilon (N-1)]^{\frac 6N}},\\
  g_1&=&-\frac{N+1}{4[\varepsilon (N-1)]^{\frac 1N}}, \\  g_2&=&\frac{N+1}{12[\varepsilon (N-1)]^{\frac 1N}}, \\
  M^{(1)}(x_s)&=&M^{(2)}(x_s)=0
\end{eqnarray*}
where $x=\frac r {r_0}$, and $\varepsilon=\frac E{\sigma S_{N-1}r_0^{N-1}}$ are the dimensionless coordinate and energy respectively. Substituting the above expressions into eq. (\ref{36}) 
yields as the condition for the first-order phase transition 
\begin{equation}
  \label{37}
  N(N-1)+\frac {19}4>0
\end{equation}
which holds for any $N$. Of course, here only the cases $N=2$ and $3$ are relevant to the tunneling and the criterion (\ref{37}) is meaningful.

\section{de Sitter Minisuperspace Model}
\label{sec:IV}

We consider the tunneling case of $N=3$ for the cosmological model of eq.(\ref{4}). Introducing the energy by $H=E$, the corresponding integrated Euclidean equation of motion reads
\begin{equation}
  \label{38}
  \frac 12 M(\xi)\dot \xi^2-V(\xi)=-E
\end{equation}
The bounce configuration $\xi_b(\tau,E)$ at finite energy is obtained and plotted in Fig. 1b. The period of the bounce at finite energy is given by
\begin{equation}
  \label{39}
  \beta(E)=\frac 4\Omega \sqrt{\frac{\xi_+}{\xi_-+\xi_i}}\left[ (1+\frac{\xi_i}{\xi_+})\Pi(\alpha^2,k)-\frac{\xi_i}{\xi_+} {\cal K}(k)\right]
\end{equation}
where $\Pi(\alpha^2,k)$ denotes the complete elliptic integral of the third kind with modulus
\begin{equation}
  \label{40}
  \alpha^2=\frac{\xi_--\xi_+}{\xi_-+\xi_+}<0, \qquad k^2=\frac{(\xi_+-\xi_-)\xi_i}{\xi_+(\xi_-+\xi_i)}
\end{equation}
where $\xi_+$, $\xi_-$ and $-\xi_i$ are the roots of the algebraic equation
\begin{equation}
  \label{41}
  \xi^3-\frac{1}{\Omega}\xi+\frac{E}{\eta}=0
\end{equation}
$\xi_\pm$ are the two turning points shown in Fig. 1, and $\xi_i \geq \xi_+ \geq \xi_-$. The period $\beta(E)$ again increases monotonically  with energy (as shown in Fig. 3) from $\beta_0=\frac{\pi}{\Omega}$ for $E=0$ to the sphaleron period $\beta_s=\frac{2\pi}{\sqrt{3}\Omega}$ at energy $E=U_0=\frac{m_0}{3^{3/2}\Omega}$. The sphaleron is
\begin{equation}
  \label{42}
  \xi_s=\frac 1{\sqrt{3} \Omega}
\end{equation}
The numerically evaluated action of the periodic bounce is shown in Fig. 4 where $S_{th}=\frac{U_0}T$ denotes the thermal action. Since the thermal action is lower than that of the bounce, the creation rate is dominated by the thermal activation over the barrier similar to the case of bubble nucleation\cite{Garriga}. It is also evident that the shallow barriers (Fig. 1 and 2) favor the thermal activation. We now turn to the crossover from quantum tunneling to thermal activation. From the equation of motion (\ref{38}) we find
\begin{eqnarray*}
  M^{(1)}&=&m_0, \qquad  M^{(2)}=0, \qquad \omega_s=\frac{2\pi}{\beta_s}= \sqrt{3}\Omega,\\
  V^{(1)}(r_s)&=&0, \qquad V^{(2)}(r_s)=-\sqrt{6 \eta m_0}, \qquad V^{(3)}(r_s)=6\eta, \qquad V^{(4)}(r_s)=0, \\
  g_1&=&0, \qquad g_2=-\sqrt{\frac{\eta}{6m_0}}
\end{eqnarray*}
With the above data the condition for the first-order phase transition (\ref{36}) becomes
\begin{equation}
  \label{43}
  -\eta\sqrt{6\eta}{m_0}<0
\end{equation}
which holds always since $\eta>0$. The phase transition is therefore of first-order, i.e. the same as that in the case of bubble nucleation.\\
The periodic bounce for the minisuperspace model at hand does not possess a bifurcation (see Fig. 4) similar to the $O(3)\sigma$-model in the 
eletroweak theory\cite{Kuznetsov,Park} 
and is
different from that of well studied spin tunneling where the sharp first order phase transition is a necessary result of bifurcation in the plot of instanton action
versus period. To see the crossover behaviour clearly we look at the thermal rate $\Gamma (T)$ which is constructed from P(E) by averaging with the Boltzmann exponential at temperature  T
and so equals
\begin{equation}
\label{44}
\Gamma (T)=\int_{0}^{\infty}dEe^{-\frac{E}{T}}P(E)\sim\int_{0}^{\infty}dEe^{-W(E)-\frac{E}{T}}
\end{equation}
In the weak coupling limit the integral over energy E can be calculated by the steepest descent method. Only periodic instantons with the period equal to inverse temperature
can dominate the thermal rate. This is called the saddle point condition:
\begin{equation}
\label{45}
\beta(E)=\frac{1}{T}.
\end{equation}
In our case the saddle point given by eq.(45) is a minimum. The action increases monotonically when the period changes from $\beta(E=0)$ to the sphaleron period $\beta_{s}$ (see Fig.4). The
curve $S_{b}(\beta)$ is convex downward and the thermal rate eq.(44) is, therefore, saturated either by $E=0$ or by $E=U_{0}$ depending on temperature,
\begin{eqnarray*}
\Gamma\sim e^{-S_{b}(E=0)},\hspace{1.0cm} T<\frac{U_{0}}{S_{b}(E=0)}\\
\Gamma\sim e^{\frac{U_{0}}{T}},\hspace{1.0cm} T>\frac{U_{0}}{S_{b}(E=0)}
\end{eqnarray*}
Fig.5 (fat line) shows the plot of $\ln\Gamma$ versus temperature T. The sharp first order phase transition is obvious, since the transition from quantum tunneling at zero
energy jumps directly to the thermal hopping. The periodic bounces with finite energy do not contribute to the thermal nucleation rate except the zero energy bounce.    

\section{Conclusions}
\label{sec:V}

The nucleation rate of the minisuperspace models is dominated either by quantum tunneling at low temperature or by thermal activation following the Arrhenius law. The 
transition of the creation rate from the quantum to the classical region is always a phase transition of the sharp first-order.
\vskip 1cm

{\bf Acknowledgment:} This work was supported by the National Natural Science Foundation of China under Grant Nos. 19677101 and 19775033. J.-Q. L. also acknowledges support by a DAAD-K.C.Wong Fellowship.

\newpage
\begin{center}
  {\bf Figure Captions:}
\end{center}
\noindent
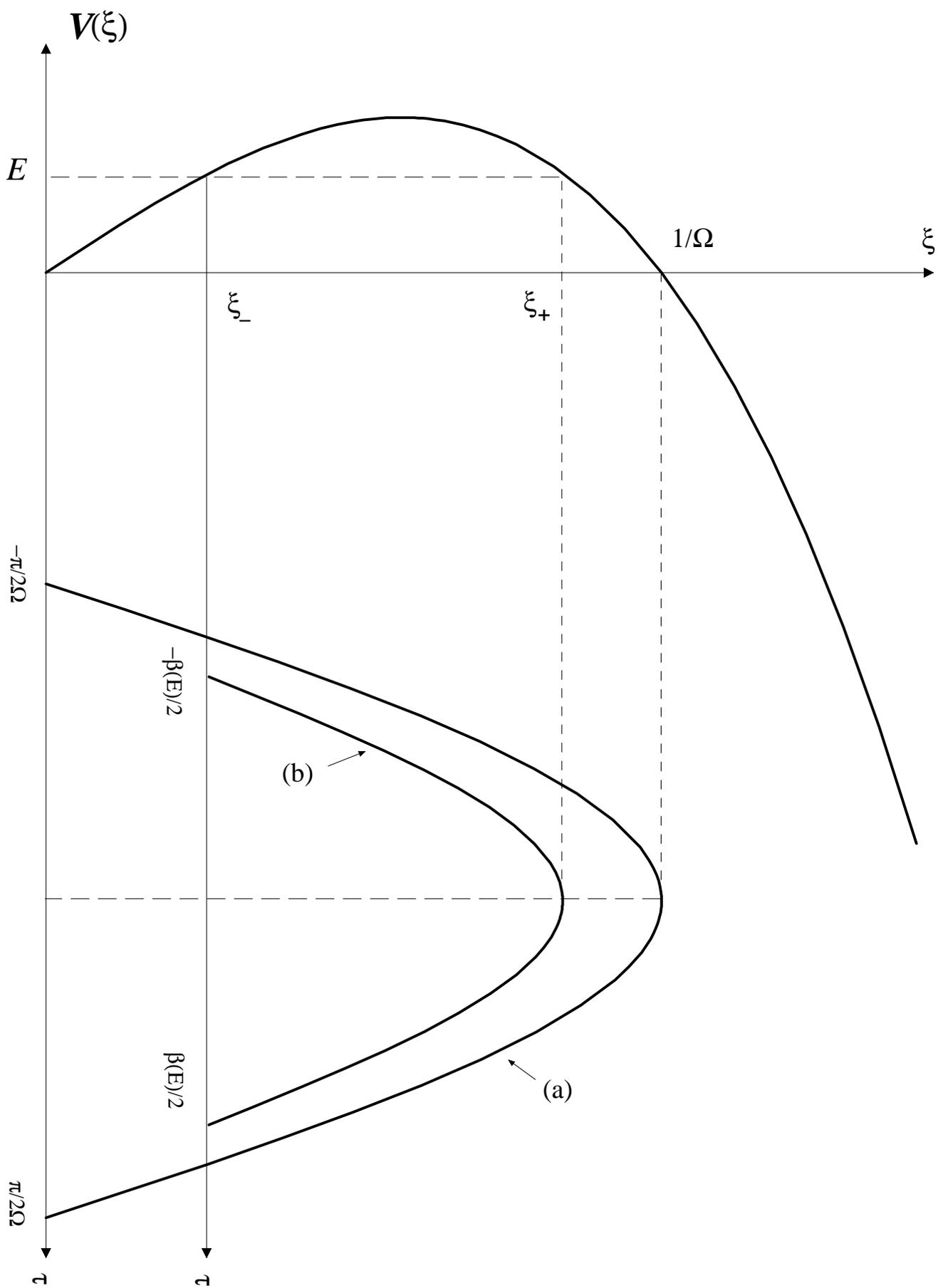
Fig. 1: The potential of eq. (\ref{6}) and the trajectories of periodic bounces: (a) The zero energy bounce and (b) the bounce at finite energy.\\
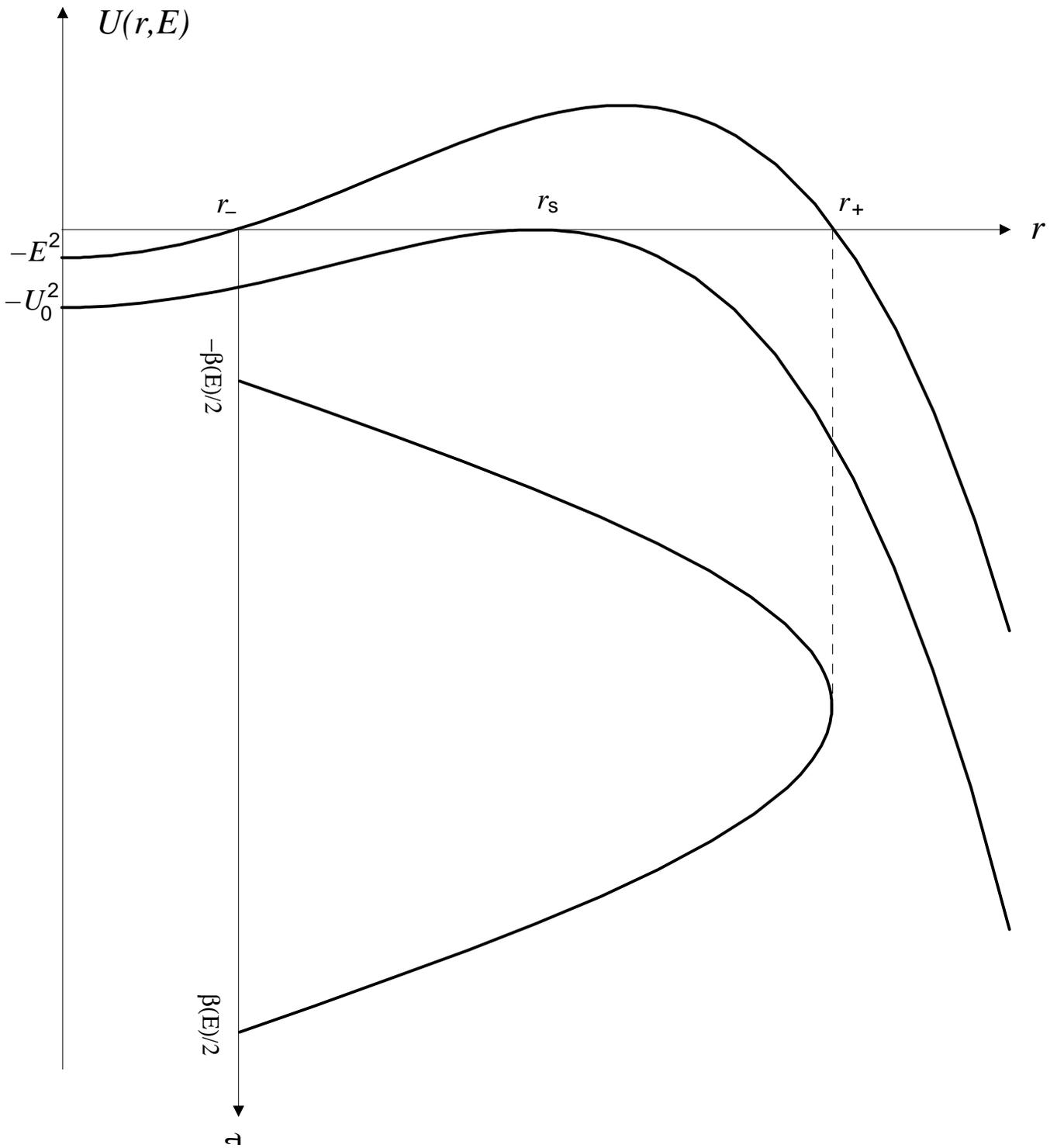
Fig. 2: The potential of eq. (\ref{26}) and the periodic bounce at finite energy $E$ .\\
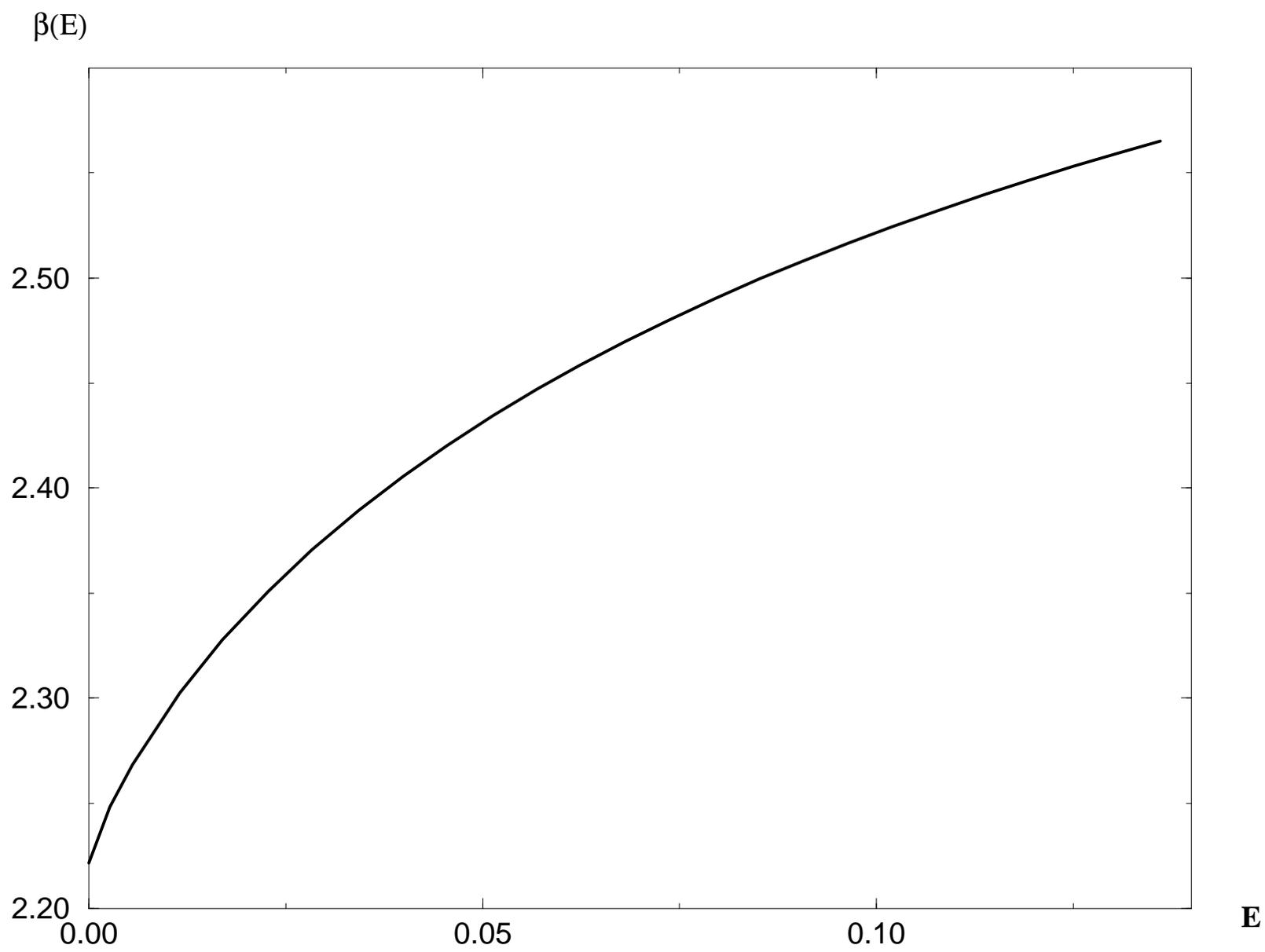
Fig. 3: The period of the bounce as a function of energy with $\eta=m_0=1$.\\
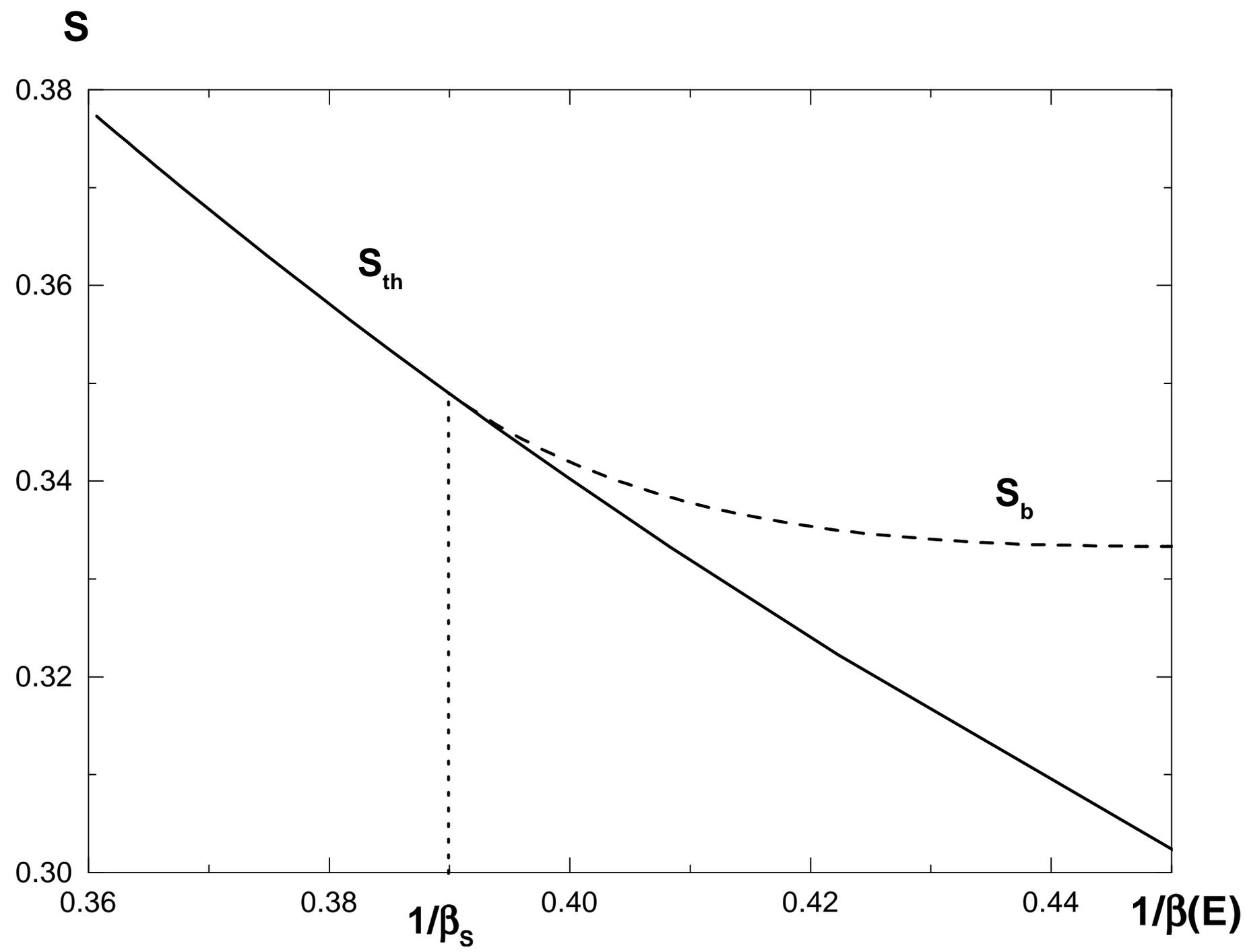
Fig. 4: The action of the bounce $S_b$ and the thermal action $S_{th}$ as functions of inverse period, with the same scale as in Fig. 3.\\
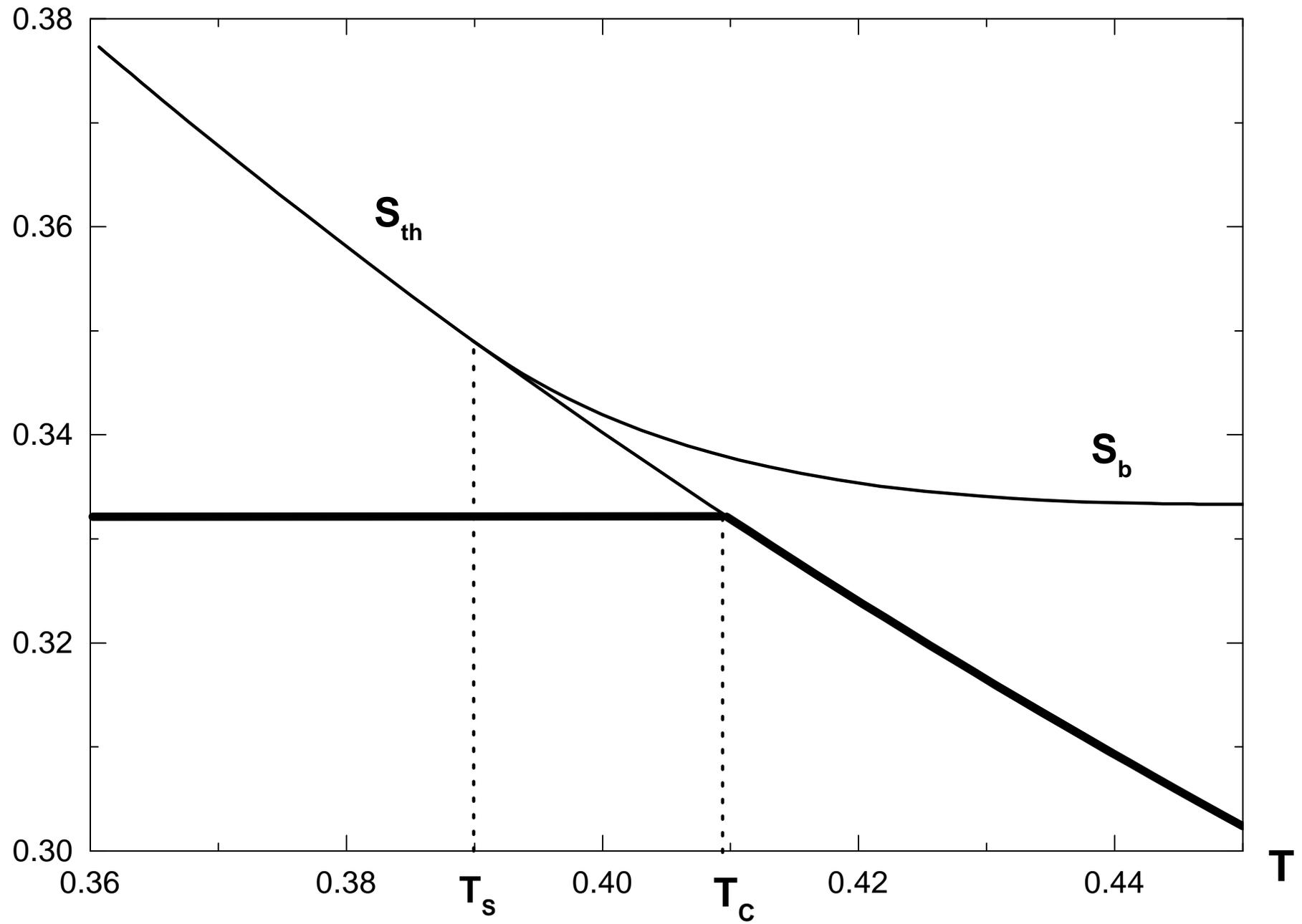
Fig. 5: The logarithmic thermal nucleation rate as a function of temperature.

\begin{thebibliography}{99}

\bibitem{Gildener}E. Gildener and A. Patrascioiu, Phys. Rev. D {\bf 16}, 423(1977).
\bibitem{Coleman}S. Coleman, Phys. Rev. D {\bf 15}, 2929 (1977); C. Callan, Jr. And S. Coleman, ibid. {\bf 16}, 1726 (1977).
\bibitem{Langer}J. S. Langer, Ann. Phys. (N. Y.) {\bf 41}, 108 (1967).
\bibitem{Liang1}J. -Q. Liang and H. J. W. M\"{u}ller-Kirsten, Phys. Rev. D {\bf 45}, 2963 (1992); ibid. {\bf 48}, 964 (1993).
\bibitem{Linde1}A. D. Linde, Phys. Lett. B {\bf 100}, 37 (1981).
\bibitem{Khlebnikov}S. Yu. Khlebnikov, V. A. Rubakov and P. G. Tinyakov, Nucl. Phys. B {\bf 367}, 334 (1991).
\bibitem{Liang2}J. -Q. Liang and H. J. W. M\"{u}ller-Kirsten, Phys. Rev. D {\bf 50}, 6519 (1994); ibid. {\bf 51}, 718 (1995).
\bibitem{Manton}N.S. Manton and T.S. Samols, Phys. Lett. B {\bf 207}, 179(1988).
\bibitem{Liang3}J.-Q. Liang, H. J. W. M\"{u}ller-Kirsten, and D. H. Tchrakian, Phys. Lett. B {\bf 282}, 105 (1992).
\bibitem{Affleck} I. Affleck, Phys. Rev. Lett. {\bf 46}, 388(1981).
\bibitem{Chudnovsky1} E. M. Chudnovsky, Phys. Rev. A {\bf 46}, 8011(1992).
\bibitem{Chudnovsky2} E. M. Chudnovsky and D. A. Garanin, Phys. Rev. Lett. {\bf 79}, 4469(1997).
\bibitem{Liang4} J.-Q. Liang, H. J. W. M\"{u}ller-Kirsten, D. K. Park, and F. Zimmerschied, Phys. Rev. Lett. {\bf 81}, 216(1998).
\bibitem{Lee} S. Y. Lee, H. J. W. M\"{u}ller-Kirsten, D. K. Park and F. Zimmerschied, Phys. Rev. B {\bf 58}, 5554 (1998)
\bibitem{Garriga}J. Garriga, Phys. Rev. D {\bf 49}, 5497(1994).
\bibitem{Ferrera}A. Ferrera, Phys. Rev. D {\bf 52}, 6717(1995).
\bibitem{Linde2}A. D. Linde, Nucl. Phys. B {\bf 216}, 421 (1983).
\bibitem{Vilenkin}A. Vilenkin, Phys. Rev. D {\bf 50}, 2581 (1994); ibid. {\bf 37}, 888 (1988); ibid. {\bf 33}, 3560 (1986).
\bibitem{Kuznetsov}A. N. Kuznetsov and P. G. Tinyakov, Phys. Lett. B {\bf 406}, 76 (1997).
\bibitem{Gorokhov} D. A. Gorokhov and G. Blatter, Phys. Rev. B {\bf 56}, 3130(1997);B {\bf 57}, 3586(1998).
\bibitem{Muller-Kirsten} H. J. W. M\"{u}ller-Kirsten, D. K. Park and J. M. S. Rana, cond-mat/9902184, to appear in Phys. Rev. B.
\bibitem{Park} D. K. Park, Hungsoo Kim and Soo-Young Lee, hep-th/9904047, to appear in Nucl. Phys. B.


\end{thebibliography}
\end{document}